# VHDL Modeling of Intrusion Detection & Prevention System (IDPS) – A Neural Network Approach

Tanusree Chatterjee
Department of Computer Science
Regent Education and Research Foundation

Abhishek Bhattacharya
Department of Computer Science
Institute of Engineering & Management

***Abstract-*** *The rapid development and expansion of World Wide Web and network systems have changed the computing world in the last decade and also equipped the intruders and hackers with new facilities for their destructive purposes. The cost of temporary or permanent damages caused by unauthorized access of the intruders to computer systems has urged different organizations to increasingly implement various systems to monitor data flow in their network. The systems are generally known as Intrusion Detection System (**IDS**).Our objective is to implement an artificial network approach to the design of intrusion detection and prevention system and finally convert the designed model to a VHDL (**V**ery **H**igh Speed Integrated Circuit **H**ardware **D**escriptive **L**anguage) code. This feature enables the system to suggest proper actions against possible attacks. The promising results of the present study show the potential applicability of ANNs for developing practical IDSs.*

## I. INTRODUCTION

An **IDS** can be a piece of installed software or a physical appliance that monitors network traffic in order to detect unwanted activity and events such as illegal and malicious traffic, traffic that violates security policy and that violates acceptable use policies. Intrusion Detection Systems (IDS) are now mainly employed to secure company networks. With the explosive growth of the network systems, information exchange became routine between computers around the world, thus the need for network security has become even more critical with the rise of information technology in everyday life. Meanwhile, the complexity of attacks is on the rise regardless of the beefed up security measures. Intrusion Prevention System (**IPS**) provides an in-line mechanism focus on identifying and blocking malicious network activity in real time. An IPS is a type of IDS that can prevent or stop unwanted traffic. The IPS usually logs such events and related information.

The IPS is unique in its ability to gather evidence of an attacker's activity, remove the attacker's access to the network, and reconfigure the network to resist the attacker's penetration technique. The IPS stops attacks at the source of the threat and can proactively protect against future threats and vulnerabilities. In our present study, an intrusion detection system is implemented using Multi Layer Perceptron (MLP) ANN. The universal approximation theorem states that an MLP (with one or more hidden layers) can approximate any function with arbitrary precision and of course the price is an increase in the number of neurons in the hidden layer. Using more than one layer may lead to more efficient approximation or to achieving the same accuracy with fewer neurons in the neural network.

One of the objectives of the present study is to evaluate the possibility of achieving the same results with this less complicated neural network structure. Using a less complicated neural network is more computationally efficient. Also it would decrease the training time. While in some previous studies the implemented system is a neural network system with the capability of detecting normal or attack connections, in our present study a more general problem is considered in which the attack type is also detected. This feature enables the system to suggest proper actions against possible attacks. The promising results of the present study show the potential applicability of ANNs for developing practical IDSs.

KDD 99 dataset are used as the input vectors for training and validation of our neural network. It was created based on the DARPA (Defense Advanced Research Project Agency) intrusion detection evaluation program which is publicly accessible via MIT Lincoln Lab[12][13].

## II. RELATED WORK

There have been lots of traditional rule-based works to the design of IDS. Recently, an increasing amount of research has been conducted on applying artificial neural networks to detect intrusions in the network [3] [4]. This method proves to be advantageous as it goes through rigorous training, validation and testing phases before being actually fed to the network for detecting attacks. Comparing different neural network classifiers, the back-propagation neural network (BPN) has showed to be more efficient in developing IDS. In recent years, various approaches on Biological Inspired Intrusion Prevention and Self-healing System for network security have been proposed. These approaches are based upon data inspired by the human immune system (HIS) which applied to the autonomous defense system. The system integrates an artificial immune intrusion prevention system for network security inspired by the immunology theory known as danger theory and adaptive immune system. Attackers take the advantage of CGI (Common Gateway Interface) scripts to perform an attack by sending illegitimate inputs to the web server. A thesis in recent year also contains the findings and the results of the CGI related web server attacks and an explanation of the design and implementation of Intrusion Detection and Prevention System for CGI based web server attacks. In recent years, multilayer intrusion detection and prevention system architecture for VoIP infrastructures has been proposed. The key components of the approach are based on a VoIP-specific honey pot and on an application layer event correlation engine [5].Some research in recent years has been done for IDPS using object oriented analysis method which describes the state's overall requirements regarding the acquisition and implementation of IDPS with intelligence.

## III. PROCEDURAL DETAILS

We have started our work towards designing an IDPS using Neural Network in MATLAB 2010b. We have used all 41 attributes of the KDD 99 dataset to train the ANN and then supplying it with other data sets which are yet to be classified to what type of attack it is. Data being used are live data from MIT Labs.

We started with the following activities before going to build the SIMULINK model of IDPS.
- Data collection
- Data Cleaning and Data Pre-Processing
- Classification of different attacks
- Data Feeding to ANN
- Use of basic nprtool
- Simulating Results.
- Observing Results.

**Data Collection**
KDD 99 data set based on the DARPA intrusion detection program, which is publicly accessible via MIT Lincoln Lab, is first of all collected at this block. We started our work with 5000 data of the total dataset, after that 25000 and then we proceeded with 1, 00000 and 3, 00000 dataset.

**Data Cleaning and Data Pre-Processing**
After collecting the original data from the MIT Lincoln Lab, we extracted the required features, and converted the data set into Matlab compatible format. This basically performs the data cleaning and pre processing procedure. First we clear the total dataset – we have numbered each element of the all 42 attributes. Then we converted all the elements to double and classified all the attacks according to the number resided in 42 th column.

The attributes given in the data set are converted into double data type to make it compatible with the ANN Tool box of Matlab. Then the feature variables have been converted e.g. the "protocol type" with values like tcp=0, udp=1, icmp=2. The attacks in the data set have been categorized as DOS=1, normal=0, probe=2, r2l=3, u2r=4.

Previous work has been done in the field of detection using limited number of attributes of KDD Dataset 99. We have used all 41 attributes (shown in figure 1.1) of the dataset to train the ANN and then supply it with other unclassified data sets which are yet to be classified.

| Feature | Description | Type | Feature | Description | Type |
|---|---|---|---|---|---|
| 1. duration | Duration of the connection. | Cont. | 22. is guest login | 1 if the login is a "guest" login; 0 otherwise | Disc. |
| 2. protocol type | Connection protocol (e.g. tcp, udp) | Disc. | 23. Count | number of connections to the same host as the current connection in the past two seconds | Cont. |
| 3. service | Destination service (e.g. telnet, ftp) | Disc. | 24. srv count | number of connections to the same service as the current connection in the past two seconds | Cont. |
| 4. flag | Status flag of the connection | Disc. | 25. serror rate | % of connections that have "SYN" errors | Cont. |
| 5. source bytes | Bytes sent from source to destination | Cont. | 26. srv serror rate | % of connections that have "SYN" errors | Cont. |
| 6. destination bytes | Bytes sent from destination to source | Cont. | 27. rerror rate | % of connections that have "REJ" errors | Cont. |
| 7. land | 1 if connection is from/to the same host/port; 0 otherwise | Disc. | 28. srv rerror rate | % of connections that have "REJ" errors | Cont. |
| 8. wrong fragment | number of wrong fragments | Cont. | 29. same srv rate | % of connections to the same service | Cont. |
| 9. urgent | number of urgent packets | Cont. | 30. diff srv rate | % of connections to different services | Cont. |
| 10. hot | number of "hot" indicators | Cont. | 31. srv diff host rate | % of connections to different hosts | Cont. |
| 11. failed logins | number of failed logins | Cont. | 32. dst host count | count of connections having the same destination host | Cont. |
| 12. logged in | 1 if successfully logged in; 0 otherwise | Disc. | 33. dst host srv count | count of connections having the same destination host and using the same service | Cont. |
| 13. # compromised | number of "compromised" conditions | Cont. | 34. dst host same srv rate | % of connections having the same destination host and using the same service | Cont. |
| 14. root shell | 1 if root shell is obtained; 0 otherwise | Cont. | 35. dst host diff srv rate | % of different services on the current host | Cont. |
| 15. su attempted | 1 if "su root" command attempted; 0 otherwise | Cont. | 36. dst host same src port rate | % of connections to the current host having the same src port | Cont. |
| 16. # root | number of "root" accesses | Cont. | 37. dst host srv diff host rate | % of connections to the same service coming from different hosts | Cont. |
| 17. # file creations | number of file creation operations | Cont. | 38. dst host serror rate | % of connections to the current host that have an S0 error | Cont. |
| 18. # shells | number of shell prompts | Cont. | 39. dst host srv serror rate | % of connections to the current host and specified service that have an S0 error | Cont. |
| 19. # access files | number of operations on access control files | Cont. | 40. dst host rerror rate | % of connections to the current host that have an RST error | Cont. |
| 20. # outbound cmds | number of outbound commands in an ftp session | Cont. | 41. dst host srv rerror rate | % of connections to the current host and specified service that have an RST error | Cont. |
| 21. is hot login | 1 if the login belongs to the "hot" list; 0 otherwise | Disc. | | | |

**Figure 1.1 All 41 features of KDDCUP'99 dataset**
**Classification of different attacks**

To classify the different attacks we have found out the four main attack categories (Table 1.1):

a. **DOS (DENIAL OF SERVICE): AN ATTACKER TRIES TO PREVENT LEGITIMATE USERS FROM USING A SERVICE E.G. TCP SYN FLOOD, SMURF ETC.**
b. **Probe:** An attacker tries to find information about the target host e.g. scanning the target host in order to get information about available resources.
c. **2R (User to Root):** An attacker has local account on victim's host and tries to get the root's privileges.
d. **R2L (Remote to Local):** An attacker does not have local account on the victim's machine and tries to obtain it.

**And another category we have mentioned as**
e. **Other:** Those which are yet to be classified.

So we categorized the attacks in 5 categories and numbered them as – Dos (1), Probe (2), U2R (3), R2L (4) and other (5). We have numbered the normal data as 0.

| DoS | U2R | Probe | R2L | Other |
|---|---|---|---|---|
| Back | Buffer overflow | lpsweep | FTP write | Named |
| Land | Load module | Nmap | Guess passud | Xlock |
| Neptune | Perl rootkit | Portsweep | | |
| Pod | | Satan | lmap | |
| Smurf Teardrop | | | Multihop Physpy Warezclient Warezmaster | |

**Table 1.1 Different attacks of KDD dataset**

## IV. EXPERIMENTAL RESULTS

Simulation was performed in MATLAB using dataset containing 311030 sample data was used as input to the IDS. Out of this 217720 samples were used for training, 46655 samples were used for validation and the rest 46655 samples were used for testing.

The performance of the network taking into account training, validation and testing data in Matlab for the given dataset (using all the 41 features) is shown in the following figures.

Total number of epochs taken is 243 where the validation check is 6, at epoch 243 and there is no validation failure up to this epoch (shown in figure 1.2).

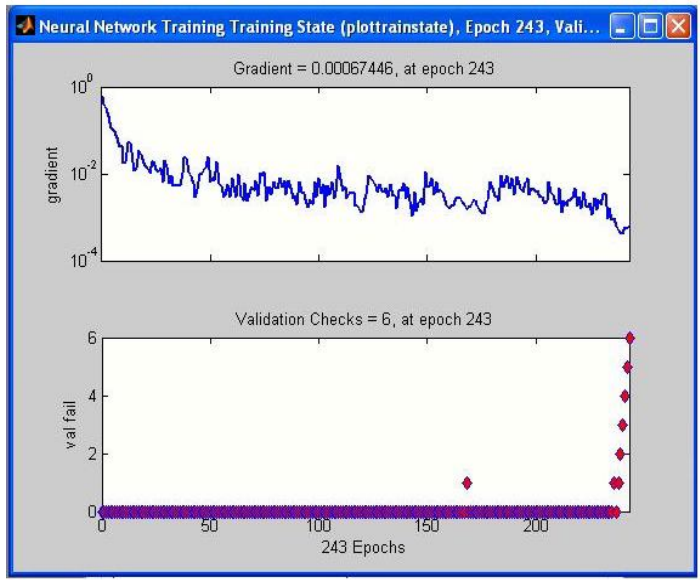

**Figure 1.2 Neural Network Training state and Validation check graph**

We have obtained a least mean square error of $10^{-2}$ and the best validation performance was found to be 0.0088598 at epoch 237(Figure 1.3).

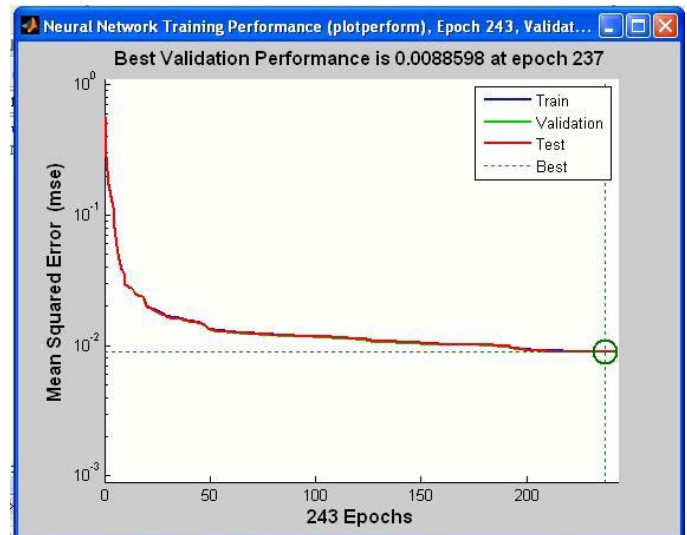

**Figure 1.3 NN Performance graph**

The next figure (figure 1.4) shows the confusion matrices for training, testing, and validation, and the three kinds of data combined. The network outputs are very accurate, as you can see by the high numbers of correct responses in the green squares and the low numbers of incorrect responses in the red squares. The lower right blue squares illustrate the overall accuracies. It shows the Success Rate vs. Error Rate in all the stages like training, validation and testing.

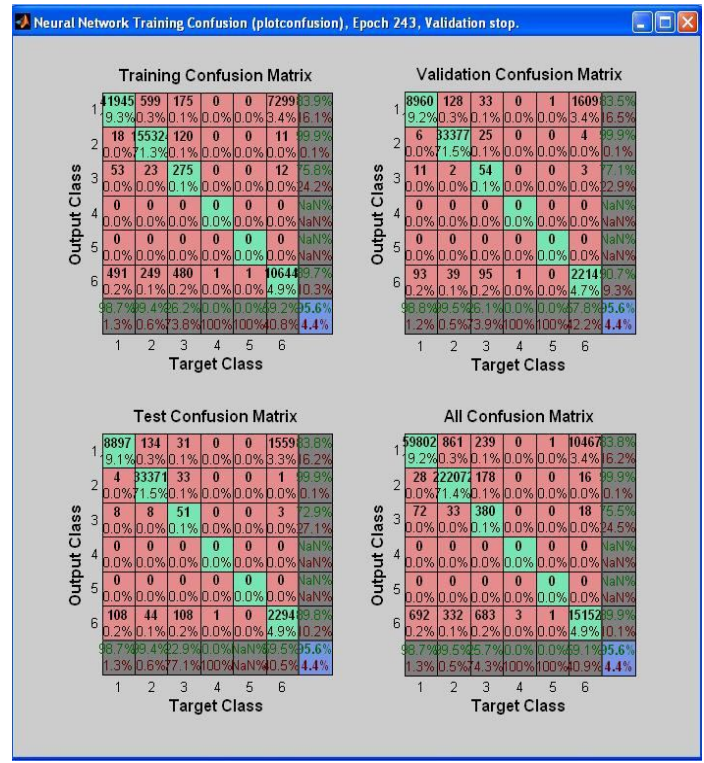

**Figure 1.4 Confusion Matrix**

The following curves (Figure 1.5) represent the receiver operating curve. The colored lines in each axis represent the ROC curves.

The *ROC curve* is a plot of the true positive rate (sensitivity) versus the false positive rate (1 - specificity) as the threshold is varied. A perfect test would show points in the upper-left corner, with 100% sensitivity and 100% specificity. For this problem, the network performs very well.

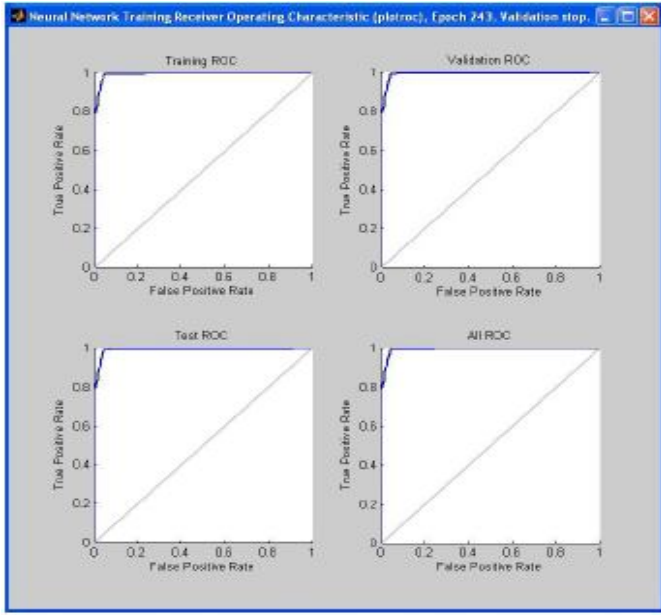

**Figure 1.5 Receiver operating characteristic graph – true positive vs. false positive**

## True and False alarms generated by IDS

- **True Positive:** A legitimate attack which triggers an IDS to produce an alarm.

- **False Positive:** An event signaling an IDS to produce an alarm when no attack has taken place.

- **False Negative:** A failure of an IDS to detect an actual attack.

- **True Negative:** When no attack has taken place and no alarm is raised.

**The results are enumerated in the following table.**

| Phase | Success rate | Failure rate |
|---|---|---|
| Training | 95.6% | 4.4% |
| Validation | 95.6% | 4.4% |
| Testing | 95.6% | 4.4% |

**Table 1.2 Success and failure rate of all stages**

### v. CONCLUSION AND FUTURE SCOPE

Our main objective is to convert the above design of the simulink model of the IDPS to VHDL code and downloaded next to FPGA (Sparatn 3 Starter Board). Now we are trying to convert the model of single layer ANN attack classifier model and next we will do the same using the multilayer structure described above. After converting the classifier model which will classify the normal or attack data separately, we will implement the VHDL of the total IDPS model which will not only detect the attack but also prevent them. Now we are working in offline scenario, after the successful completion of our work we will implement our architecture in online scenario.

**AUTHORS PROFILE**

**First Author -**Tanusree Chatterjee, M.Tech (CSE) from WBUT, Assistant Professor in Department of Computer Science, Regent Education and Research Foundation, Barrackpore, West Bengal.

**Second Author** –Abhishek Bhattacharya, M.Tech (CSE) from BIT,Mesra, Assistant Professor in Department of Computer Science, Institute of Engineering & Management, Saltlake, Kolkata, West Bengal.